\documentclass[letterpaper,12pt]{article}   
\usepackage{osajnl2} 
\usepackage[draft]{hyperref} 
\usepackage{amssymb}		
\usepackage{epstopdf}

\begin{document}


\title{Enhancing the spectral response of filled bolometer arrays for submillimeter astronomy}


\author{Vincent Rev\'eret,$^{1,*}$ Louis Rodriguez,$^1$ and Patrick Agn\`ese$^{2}$}
\address{$^1$Laboratoire AIM Paris-Saclay, CNRS/INSU CEA/Irfu Universit\'e Paris-Diderot, F-91191 Gif-sur-Yvette, France}
\address{$^2$CEA Grenoble, LETI/SLIR, MINATEC, 17 rue des Martyrs, 38054 Grenoble, France}

\address{$^*$Corresponding author: vincent.reveret@cea.fr}

\begin{abstract}
The future missions for astrophysical studies in the submillimeter region will need detectors with very high sensitivity and large field of view. Bolometer arrays can fulfill these requirements over a very broad band.  We describe a technique that enables bolometer arrays that use quarter wave cavities to have a high spectral response over most of the submillimeter band. This technique is based on the addition on the front of the array of an anti-reflecting dielectric layer. The optimum parameters (layer thickness and distance to the array) are determined by a 2D analytic code. This general principle is applied to the case of the Herschel PACS bolometers (optimized for the 60 $\mu$m - 210 $\mu$m band). As an example, we demonstrate experimentally that a PACS array covered by a 138 $\mu$m thick silicon layer can improve the spectral response by a factor 1.7 in the 450 $\mu$m band.

\end{abstract}

\ocis{040.0040, 040.1240, 040.2235, 350.1270.}


\maketitle 
\section{Introduction}

The recently launched Herschel Space Observatory has provided astronomers with a spectacular and promising glimpse of the submillimeter universe \cite{Pilbratt}. For the next generation submillimeter instruments operating from space \cite{Nakagawa} or from ground-based facilities \cite{{Guesten},{Sebring}}, the improvement of performance generally implies to have better detector sensitivities and larger fields of view.  
Another important point is the ability to fully cover the region between far-infrared and millimeter (typically between 50 $\mu$m and 2 mm). This means that the spectral response of the detector has to be optimum in the considered spectral window. 

Bolometers are the detectors of choice for direct detection in this part of the spectrum. To get a high absorption efficiency, modern bolometer arrays generally use quarter wave cavities to absorb the incoming radiation from the telescope. They are therefore restricted, by design, to work in a specific spectral window. It is not a straightforward process to modify  the spectral range of operation of this type of detectors once assembled, because their cavities are usually made by spacers, etched materials or mechanical handling systems. However, when groups provide detectors to a project, many models are produced and only the ones with the perfectly matching parameters are selected for the mission. But some non-selected detectors can potentially give good performances in other bands, if their spectral response is enhanced. We have studied   a technical solution which is simple, fast to assemble and not expensive, that enables a detector originally designed for a single band to work in other spectral windows. The solution is based on an anti-reflecting (AR) layer whose optimum parameters are obtained by modeling.

This paper first describes the model used to obtain the spectral response:  a 2D analytic method which is fast, easy to implement  and more flexible than heavy 3D electromagnetic software.  Then the method is applied to the particular case of Herschel PACS bolometers.  Our group has developed these semiconductor sensors for the PACS photometer  \cite{{Poglitsch},{Billot06}}. These bolometers are optimized for the 60 $\mu$m - 210 $\mu$m band, but their spectral response decreases toward longer wavelengths (see next section).  With the new method described in the paper, we have found that simple configurations  with AR layers enable the PACS detectors to operate at longer wavelengths  (bands beyond 300 $\mu$m).  We focused our development on the 450 $\mu$m atmospheric window which is accessible from ground-based telescopes and which is of particular interest for astrophysical science. The results of simulation are compared to spectrometer measurements for validation in section 3.


\section{Modeling the Bolometer Spectral Response}

\subsection{Absorption principle of bolometers with $\lambda/4$ cavities. Case of the PACS bolometers}

Bolometer camera instruments for far-infrared to millimeter astronomy traditionally use integrating cavities and circular-aperture feedhorns  to couple  the detector to the telescope beam \cite{Richards94}. Generally, $2F\lambda$ or $1F\lambda$ horns were used to populate the focal planes of submillimeter bolometer cameras, with the disadvantage of undersampling the image. This is the main reason why many submillimeter cameras now use arrays of $0.5F\lambda$  bare detectors that instantaneously fully sample the image \cite{Griffin02}, and in most of the cases, the absorption is made by using $\lambda/4$ cavities.

\begin{figure}[htbp]
  \centering
  \includegraphics[width=4.5cm, angle=270]{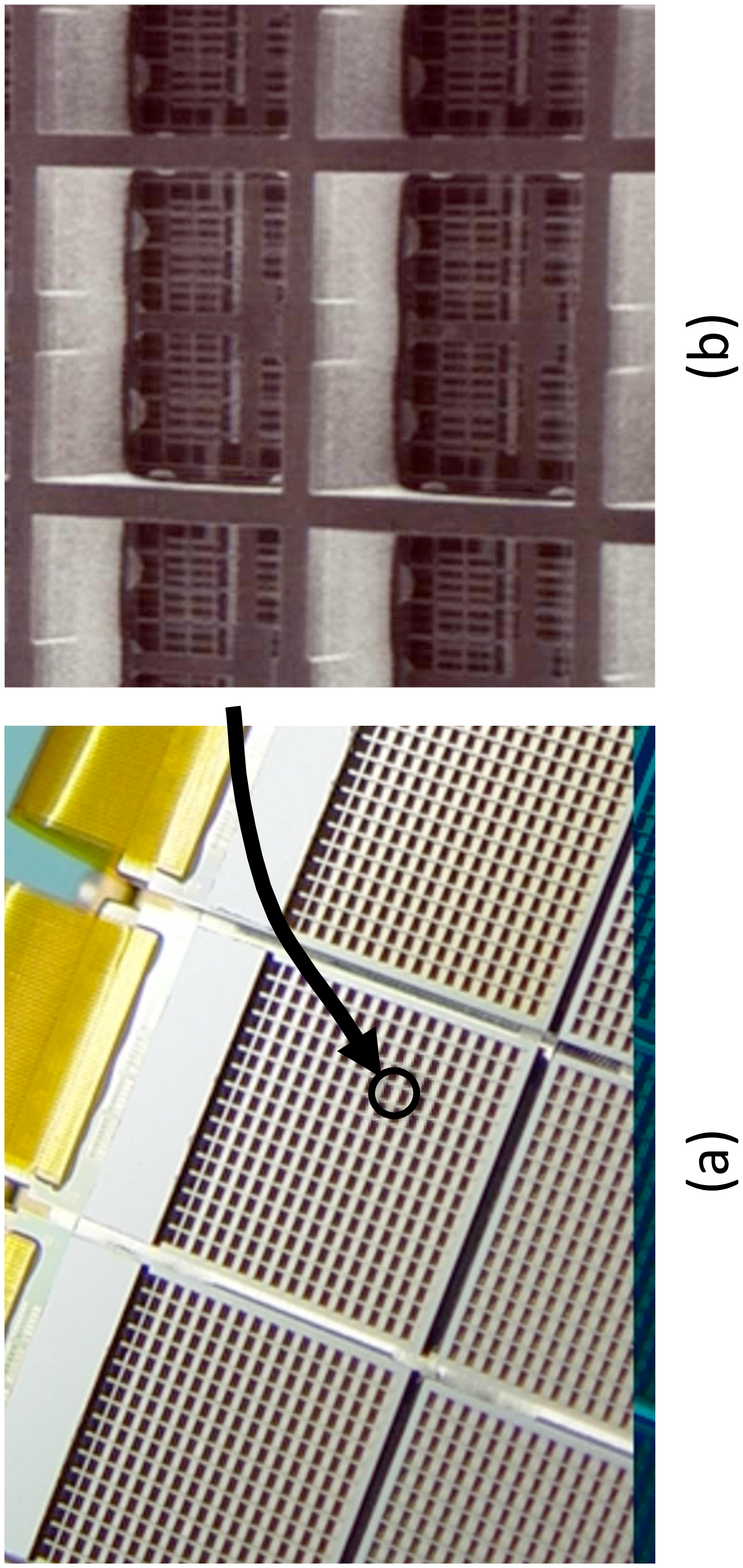}
  \caption{(Color online) (a) View of the PACS $16 \times 16$ bolometer arrays. (b) Microscope view of single bolometers inside an array.  }
  \label{Figure1}
  \end{figure}

We now take the example of the Herschel PACS detectors to describe the general behaviour of bolometers with such cavities. The bolometers that we have developed for Herschel were the first ones to use this cartesian architecture for far-infrared / submillimeter astronomy (see  Fig. \ref{Figure1}, \cite{Agnese99}). The principle of wave absorption is based on the Dennison-Hadley interference filters \cite{Hadley47}, sometimes called Salisbury screens \cite{{Salisbury},{Fante88}}. This consists in a reflecting metal layer placed at the bottom of a cavity (backshort), normal to the plane of propagation of the wave. In such a device, because of interferences between the incident and the 180$^{\circ}$  phase shifted reflected waves, a standing wave develops with a minimum of electric field intensity at the backshort and a maximum at a distance $\lambda/4$ above it. If an absorber is placed there (typically a resistive layer with a square resistance equal to the impedance of free space), 100$\%$ of the energy at this wavelength $\lambda$ can be absorbed. This absorbed incident power is converted to thermal energy and a thermometer measures the temperature rise.

\begin{figure}[htbp]
  \centering
  \includegraphics[width=6cm, angle=270]{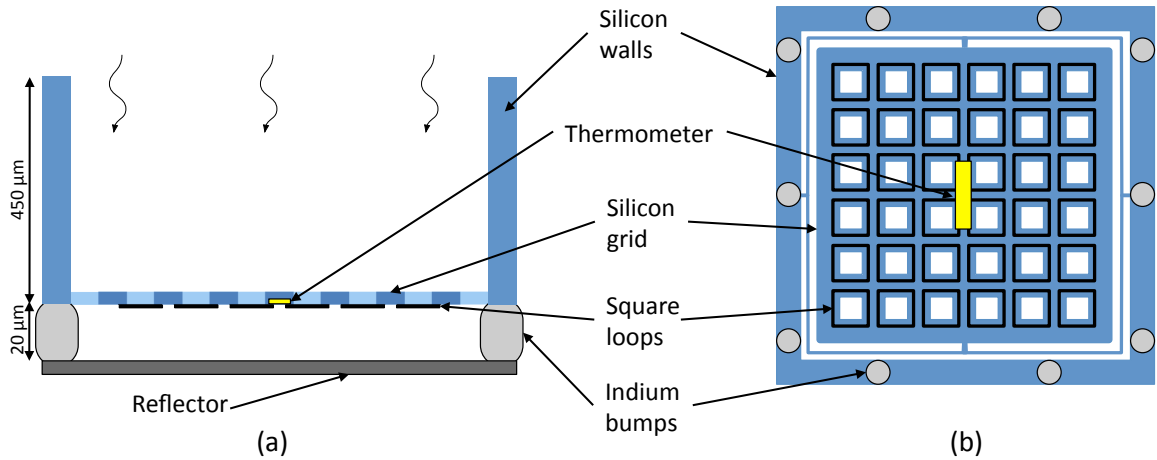}
  \caption{(Color online) (a) Side view of the quarter-wave cavity architecture (Salisbury screen) of the PACS bolometric pixel. The absorber (square loops) is deposited on the back side of the silicon grid. (b) View of the pixel grid from below. The Si:P:B thermometer is located at the center of the silicon grid. The grid is linked to the interpixel walls via narrow silicon beams (2 $\mu$m wide). The silicon walls are 450 $\mu$m high and 70 $\mu$m thick. The pixel pitch is 750 $\mu$m.   }
  \label{Figure2}
  \end{figure}

For PACS, we use a semiconducting thermometer (phosphorous doped silicon, compensated with boron impurities, \cite{Zhang93}) located at the center of an etched silicon membrane, working at cryogenic temperature (300 mK). The silicon film is patterned as a thin grid in order to reduce the heat capacity  and the cross-section to cosmic rays \cite{Bock95}. A very thin layer of absorber is deposited on the supporting silicon grid and  absorbs the incoming submillimeter radiation (see Fig. \ref{Figure2}). This is not a uniform film, but a periodic resonant structure made of square loops, to improve the spectral response in a certain optical band (see next section for the details). The vacuum quarter-wave cavity is obtained by  indium bumps hybridization techniques. These spacers have three distinct roles: (i) define the cavity height, (ii) make a thermal link between the silicon detection layer and the substrate which is kept at a constant temperature, (iii) conduct the electrical signals between the thermometer and the cold-stage electronics which is located near the reflector. In PACS, the indium bumps are 20 $\mu$m thick. This enables us to cover the region between 60 $\mu$m and 210 $\mu$m with two distinct focal planes and three different optical filters \cite{Poglitsch}. Ideally, the detection of longer wavelengths should be possible by increasing the size of the indium bumps, but re-hybridization with larger structures is not a straightforward process.

From Fig. \ref{Figure2}b it should be noted that the grid does not extend over the entire pixel surface (fill factor $= 85\%$). Mirotznik et al. \cite{Mirotznik} have studied the influence of the area fill factor on the absorption for infrared detectors using $\lambda/4$ cavities. They showed that because of diffraction effects on the sides of the absorbing surface, the absorption can be improved by a factor up to 3 compared to the purely geometric estimate. We applied this principle to longer wavelengths and found that for PACS pixels (750 $\mu$m pixel pitch) at 150 $\mu$m, the correction factor is 1.1 (corresponds to a pixel pitch of 50 $\mu$m at a wavelength of 10 $\mu$m in Fig. 3 in \cite{Mirotznik}).

\subsection{Characteristic matrix method}

\begin{figure}[htbp]
\centering
\includegraphics[width=4.5cm, angle=270]{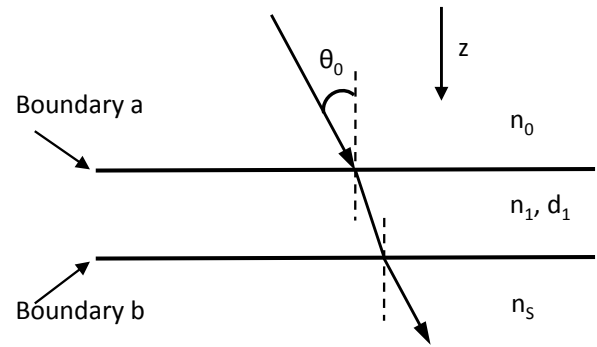}
\caption{Simple case of a thin film with refractive index $n_1$ and thickness $d_1$, over a substrate with refractive index $n_s$. }
\label{Figure3}
\end{figure}

The model used to compute the spectral response of the bolometers is based on the method of characteristic matrix. This method was first described by Abel\`es \cite{Abeles}, and is now widely used in the thin film domain thanks to its ability to be rapidly  solved by common programming languages.  Considering a thin dielectric layer of thickness $d_1$ and refraction index $n_1$, it is possible to build a $2\times2$ matrix that links the electric fields of the two interfaces $a$ and $b$ of the film, see Fig. \ref{Figure3}. 

For one layer, the relation is  (from \cite{Macleod})

\begin{equation}
\left(
\begin{array}{c}
  E_{a}   \\
  H_{a}
\end{array}
\right)=
\left(
\begin{array}{cc}
\cos\beta_1 & (i\sin\beta_1)/\eta_1\\
i\eta_1 \sin\beta_1 & \cos\beta_1
\end{array}
\right)
\left(
\begin{array}{c}
E_b\\
H_b
\end{array}
\right)
\label{matabeles}
\end{equation}
where $E_a$ is the total electric field at the first interface (sum of incident and reflected fields), $H_a$ is the total magnetic field at this same boundary. $E_b$ and $H_b$ are respectively the total electric field and magnetic field at interface $b$. The optical admittance of the medium is defined by $\eta_1=n_1\eta_0 cos\theta_0$ for TE modes and  $\eta_1=n_1\eta_0 / cos\theta_0$ for TM modes. $\eta_0=1/Z_0=1/377\Omega$  is the free space admittance and $\theta_0$ is the incident angle on the first layer. The phase shift $\beta_1$ due to the propagation through the medium  is

\begin{equation}
\beta_1=\frac{2\pi n_1d_1\cos\theta_0}{\lambda}
\end{equation}
where $\lambda$ is the wavelength in vacuum. The optical admittance is also defined by the general form
\begin{equation}
\eta=\frac{H}{E}
\label{admittance} 
\end{equation} 
The equation (\ref{matabeles}) transforms into

\begin{equation}
\left(
\begin{array}{c}
  E_{a}/E_b   \\
  H_{a}/E_b
\end{array}
\right)=
\left(
\begin{array}{c}
M\\
\end{array}
\right)
\left(
\begin{array}{c}
1\\
\eta_s
\end{array}
\right)
\label{matabeles2}
\end{equation}
where  $\left(M\right)$ is the characteristic matrix of the  layer \textit{m} and $\eta_s$ is the admittance of the exit medium (substrate). Defining $B= E_{a}/E_b$ and $C=H_{a}/E_b$, then for a stack of $q$ layers, Eq. (\ref{matabeles2}) becomes

\begin{equation}
\left(
\begin{array}{c}
  B  \\
  C
\end{array}
\right)=
\left \{
\prod_{r=1}^q
\left(
\begin{array}{cc}
\cos\beta_r & (i\sin\beta_r)/\eta_r\\
i\eta_r\sin\beta_r & \cos\beta_r
\end{array}
\right)
\right \}
\left(
\begin{array}{c}
1\\
\eta_s
\end{array}
\right)
\label{matabeles3}
\end{equation}
which is the general expression for thin films theory. The amplitude reflexion coefficient at the entrance of the multi-layer is defined by

\begin{equation}
r=\frac{\eta_0-\eta}{\eta_0+\eta}
\label{reflex1}
\end{equation}
where $\eta=C/B$ is the optical admittance of the assembly. Combining Eqs. (\ref{admittance}) and (\ref{reflex1}), the reflexion coefficient can be written as

\begin{equation}
r=\frac{B\eta_0-C}{B\eta_0+C}
\label{reflex2}
\end{equation}

The reflectance of the multi-layer stack is therefore given by $R=r\times r^{\ast}$. As we are interested in the spectral response of the detector (its absorptance $A$), the model will evaluate the quantity $A(\lambda)=1-R(\lambda)$, assuming that no radiation exits the back of the bolometer. The advantage of this method is that each layer in the stack is only characterized by a matrix and two parameters ($n$ and $d$), making it well suited for computer modeling and optimization.

\subsection{Spectral response of a simple cavity}

The matrix method is applied to the case of the PACS bolometers by considering a stack of several thin layers (see Fig. \ref{Figure2}):  a reflector, a vacuum quarter-wave cavity,  an absorber and  a supporting thin silicon grid. Except for the  quarter-wave cavity, all the other layers can not be considered as simple dielectric layers. The following sub-sections describe for each of these media the characteristic parameters that will be used in the matrix calculation. These  parameters (thickness and refractive index) are listed in Table \ref{table1}. As this method works with semi-infinite layers, the results of the simulation need to be confirmed by 3D finite-element software (see section C. 4). 

\subsubsection{Reflector}

In order to get total reflection in the submillimeter wavelength range, the skin depth inside the metal forming the reflector has to be much smaller than the physical thickness.  The skin depth is defined by  

\begin{equation}
\delta=\sqrt{\frac{2}{\omega \mu_0 \mu_r \sigma}}
\label{skindepth}
\end{equation}
where $\omega$ is the radian frequency, $\mu_0$ is the permeability of free space ($4\pi \times 10^{-7}$ H/m), $\mu_r$ is the relative permeability of the metal and $\sigma$ its electrical conductivity. For PACS, the metallic reflector consists in a 150 nm layer of gold deposited on a silicon substrate. From \cite{Matula}, the electrical conductivity for gold at  sub-Kelvin temperatures is $5 \times 10^9$ S/m, so at a wavelength of 100 $\mu m$ and with $\mu_r=1$, the skin depth is $\delta \simeq 4$ nm, which is small compared to the gold thickness used in our devices.

In the schematic view of Fig. \ref{Figure3}, this reflective layer corresponds to the substrate, with the difference that no wave is transmitted through it. At the surface of the metallic layer (the boundary $b$), the tangential electric field $E_{b}$ is equal to 0. So $E_{b}=E_{b}^+ + E_{b}^- = 0$ and from \cite{Macleod} the tangential magnetic field is $H_{b} = \eta_1 E_{b}^+ -\eta_1 E_{b}^-$ (the positive sign describes incident waves, along the $z$ axis, and the negative sign describes waves travelling in the opposite direction, $\eta_1$ is the optical admittance of medium 1, the quarter wave cavity). Combining the two equations gives $H_{b} = 2\eta_1 E_{b}^+$. By replacing $E_{b}$ and $H_{b}$ by these new values in Eq. (\ref{matabeles}), the Eq. (\ref{matabeles2}) can be rewritten as

\begin{equation}
\left(
\begin{array}{c}
  E_{a}/E_b^+   \\
  H_{a}/E_b^+
\end{array}
\right)=
\left(
\begin{array}{c}
  B'   \\
  C'
\end{array}
\right)=
\left(
\begin{array}{c}
M\\
\end{array}
\right)
\left(
\begin{array}{c}
0\\
2\eta_1
\end{array}
\right)
\label{matabeles4}
\end{equation}
The optical admittance of the assembly in that special case is $\eta=C'/B'$, so Eq. (\ref{reflex2}) can still be used with $B'$ and $C'$ replacing $B$ and $C$ respectively.

\subsubsection{Absorber}
\begin{figure}[htbp]
\centering
\includegraphics[width=5cm, angle=270]{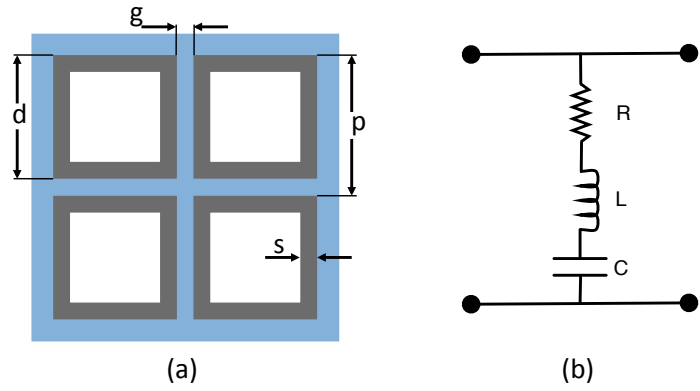}
\caption{(Color online) (a) Sketch of the TiN square-loop array deposited on the structured silicon grid (light gray). p = 35 $\mu m$, s = 4.4 $\mu m$, g = 4.4 $\mu m$, d = 30.6 $\mu m$. (b) The electrical equivalent (RLC circuit) of the square-loop array. }
\label{Figure4}
\end{figure}

In the infrared to submillimeter domain, absorbers are traditionally made of thin metal layers whose thickness is adjusted in order to get the appropriate impedance. In Dennison-Hadley filters, it is well known that  the maximum absorption efficiency (100\% at the peak wavelength in theory) is obtained when the impedance of the absorber matches the free space impedance \cite{Fante88}.  In the fields of radar technology and telecommunications,  absorbers with periodic resonant geometrical structures are sometimes used instead of homogenous layers, in order to increase the bandwidth (frequency selective surfaces or FSS, \cite{Langley}). The PACS bolometers were the first detectors to combine these two techniques (Dennison-Hadley filters and FSS) for submillimeter astronomy. In our design, the selected pattern is an array of square loops (Fig. \ref{Figure4}). 

Because of the resonant properties of the design, the absorber can not be considered as a simple homogenous lossy medium. Several authors have studied the electromagnetic properties of square loop FSS using equivalent circuit models \cite{{Langley}, {Monacelli}}. They developped expressions for the complex impedance, $Z_{loop}$ of square loop FSS that consist of a series RLC resonant circuit.  It is defined by

\begin{equation}
Z_{loop}=R_s+i\left(X_L+\frac{1}{B_C}\right)
\end{equation}

$R_s$  is the sheet resistance and is responsible for the resistive loss defined by $R_s= 1/\sigma_nd$, where  $\sigma_n$ is the electrical conductivity of the material and $d$ its thickness. The inductance $X_L$ comes from the finite length of the squares and the capacitance $B_C$ comes from the spacing between two loops. $X_L$ and $B_C$ are defined by

\begin{equation}
X_L=Z_0\frac{d}{p}H(p,2s,\lambda)
\end{equation}
and

\begin{equation}
B_C=4Z_0\frac{d}{p}H(p,g,\lambda)
\end{equation}

$H$ is defined in the Appendix A. For a given material and a fixed film thickness, the equivalent sheet resistance of this type of periodic structure is larger than the one of a homogenous film.  To obtain an absorber with an equivalent sheet resistance of 377 $\Omega/\square$, the material composing the loops must have a sheet resistance $R_s=377\times F_R$. In the case of square loops, this reduction factor is at first order $F_R=s/p$ (see Fig. \ref{Figure4} for the description of the geometrical parameters). Electromagnetic simulations with a finite-element software have been performed to get the maximum absorption in the PACS bands and resulted in the following results: $R_s=40 \Omega/\square$, p = 35 $\mu m$, s = 4.4 $\mu m$, g = 4.4 $\mu m$, d = 30.6 $\mu m$.

Choosing the absorbing material is a compromise between different technological constraints like the compatibility with other fabrication processes, the mechanical strength, the film deposition rate and the minimum achievable thickness. We use titanium nitride (TiN) because it satisfies most of these criteria. Given the TiN electrical conductivity at low temperature ($\sigma_n=1\times 10^6$ S/m), from $R_s=1/\sigma_n d$ we deduce the needed thickness, $d$, for impedance matching.  For TiN, $40 \Omega/\square$ correspond to a thickness of 20 nm. 

The TiN that is used for our device has a critical temperature $T_c=3.9$ K, it is therefore superconducting during the bolometer operation. Superconducting materials below their critical temperature are nevertheless resistive for AC currents induced by the incoming submillimeter wave. The BCS theory \cite{BCS} introduces the concept of complex conductivity for superconductors which is written $\sigma_s=\sigma_1-i\sigma_2$. In \cite{Mattis}, Mattis and Bardeen derived expressions for $\sigma_1/\sigma_n$ and $\sigma_2/\sigma_n$ that are function of temperature and of the electromagnetic wave frequency ($\sigma_n$ is the normal electrical conductivity of the material at $T_c$). $\sigma_1$ describes the thermally excited normal electrons, and is therefore responsible for absorption of the EM field. $\sigma_2$ represents the superconducting electrons and is ignored in our model \cite{Kautz}. To take into account the superconducting effect, instead of taking the normal sheet resistance $R_s$, we have to consider another expression for TiN:  $R^{\prime}_s=1/\sigma_s d$. In order to build the characteristic matrix of the TiN absorber, the refractive index must be obtained from the value of the complex impedance. Using the expression from \cite{Hilsum}, the complex refractive index of the thin TiN film is

\begin{equation}
\tilde{n}_{abs}=\sqrt{\frac{1}{2\rho \epsilon_0\omega}}(1-i)
\end{equation}
with  $\rho=Z_{loop}\times d$.

\subsubsection{Silicon grid}

Because the TiN layer is too thin to be self-suspended, it is  deposited on a 5 $\mu$m thick monocrystalline silicon grid that also contains the semiconducting thermometer (see Fig. \ref{Figure2}). In \cite{Motamedi}, Motamedi et al.  have estimated the effective refractive index of such subwavelength microstructures. This grid has an effective refractive index calculated by the following expression: 
\begin{equation} 
n_{eff}=\left[\frac{(1-f+fn_{Si}^2)\left[(1-f)n_{Si}^2+f\right]+n_{Si}^2}{2\left[(1-f)n_{Si}^2+f\right]}\right]^{1/2}
\end{equation}
  
where $f$ is the silicon filling factor of the grid (61$\%$), $n_{Si}$ is the refractive index of bulk silicon (3.4) and therefore $n_{eff}=2.2$.
The presence of this 5 $\mu$m silicon grid in the pixel cavity will slightly modify the spectral response compared to the simple case ``reflector / absorber''.

\subsubsection{Simuation results}

\begin{figure}[htbp]
  \centering
  \includegraphics[width=6cm, angle=270]{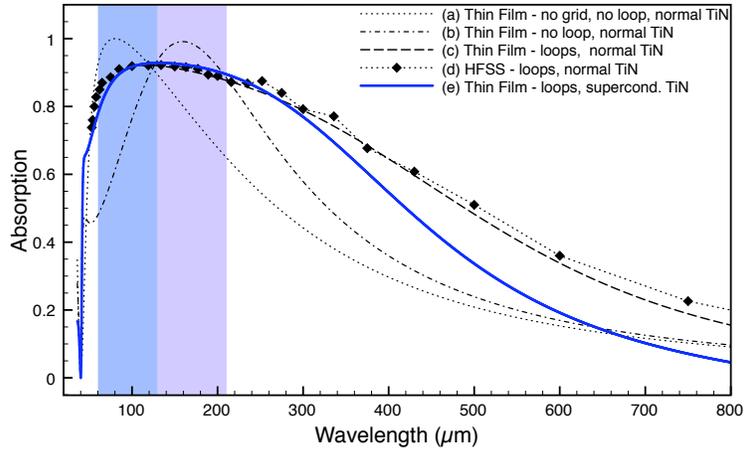}
  \caption{(Color online) Simulated absorption curves for the PACS bolometers (20 $\mu$m cavity, 35 $\mu$m square loops) for different cases (see the legend). The PACS short photometric band is represented by the blue area (60 to 110 $\mu$m) and the PACS long band is in purple (110 to 210  $\mu$m). }
  \label{Figure5}
  \end{figure}

To validate the  method of characteristic matrix as a pertinent simulation tool, the model is used to simulate the absorption of a PACS pixel (quarter-wave cavity of height $d_{c}=20$ $\mu$m and TiN loops with a pitch of  $p=35$ $\mu$m). The result is then compared to a simulation made with a finite element electromagnetic simulation software, Ansoft HFSS\texttrademark  (see Fig. \ref{Figure5}). 

The curve \textit{a} in Fig.  \ref{Figure5} corresponds to the case of a simplified cavity made of a free-standing uniform absorber (with a surface impedance of $377 \Omega/\square$, not superconducting) located 20 $\mu$m above a reflector. The spectral response is maximum at  80 $\mu$m which correponds to $\lambda=4 d_{c}$, as expected for Dennison-Hadley filters. The sudden absorption decrease toward the short wavelengths (near 40 $\mu$m) corresponds to the ``zero''  of these devices, which occurs at  $\lambda=2 d_{c}$. Adding a supporting 5 $\mu$m silicon grid above the absorber shifts the peak toward longer wavelengths (curve \textit{b}). Even if the response is very high in the case of a uniform absorber, its profile is not flat and cannot optimally cover the two PACS bands represented in blue and purple in Fig. \ref{Figure5}. This limitation is overcome by introducing the textured  absorber (curve \textit{c}).  The  absorption is relatively constant over the entire PACS band and still very high (around 90\%). The result  is very similar to the HFSS modeling (curve \textit{d}) which gives credit to the matrix method for this kind of simulation.

As it has been previously described, our model can simulate the effect of the superconductivity of TiN, for which HFSS is limited. When the effect of superconducting TiN is introduced, the long wavelengths absorption decreases compared to the case of normal TiN (curve \textit{e}). This device is limited to operation up to $\sim$ 350 $\mu$m (when the absorption drops below 70\%). This simulated absorption is compared to measurements in section 3.

\subsection{Enhancing the Response at Long Wavelengths}

\subsubsection{Demonstrator at 450 $\mu$m}

\begin{figure}[htbp]
  \centering
  \includegraphics[width=6cm, angle=270]{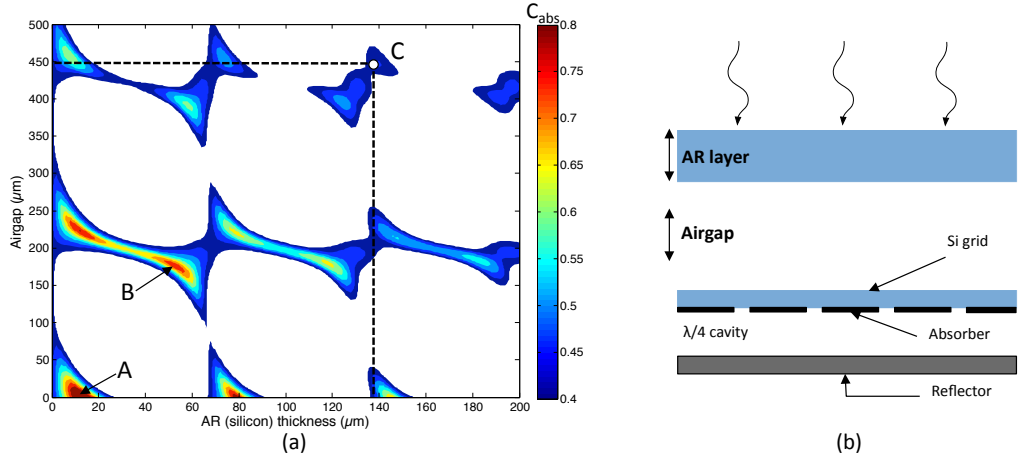}
  \caption{(Color online) (a) Average performance coefficient ($C_{abs}$) in the 450 $\mu$m band as a function of airgap thickness and silicon AR layer thickness. Cases \textit{A}, \textit{B} and \textit{C} are described in the text. (b) Thin film representation of the PACS bolometric pixel + the AR layer on the top of it (semi-infinite layers, not to scale).}
  \label{Figure6}
  \end{figure}

The matrix method is easy to modify and runs fast on a personnal computer (it takes a few seconds to obtain the curves of Fig. \ref{Figure5}). Therefore it is more adapted to test different design options than a typical finite element software. As a demonstration and example, the target band is 450 $\mu$m (425 - 470 $\mu$m), which corresponds to an atmospheric transmission band that is accessible from ground-based telescopes located on dry and high sites. Based on the idea of optical systems that use AR layers to improve reflection or transmission in a particular band, our development consists in an AR layer placed at the front-end of the bolometer array (this structure is called PACS+AR thereafter). The free parameters for the design are: (i) the distance between the dielectric and the silicon grid of the bolometer, called airgap, (ii) the thickness of the AR dielectric layer in silicon, see Fig. \ref{Figure6}b. 

The model tests the different airgap/AR combinations and estimates, for each of these cases,  the quality of the spectral response in the  450 $\mu$m band. Two criteria are evaluated to characterize the response: the average absorption coefficient in the band $\alpha$, and its flatness index $\gamma$. The flatness index is defined by the average deviation from $\alpha$  ($\gamma=100\%$ for a perfectly flat response and 0\% for a step profile). We use a single performance coefficient ($C_{abs}=\alpha \gamma$)  to find the optimum parameters that provide the best response in the band. Results are displayed in the form of a 2D map (Fig. \ref{Figure6}a). It  shows that different combinations of thicknesses give good results, but not all of them are easily feasible. Solution \textit{A} for example, which gives the best performance, corresponds to a very thin AR layer (13 $\mu$m) located just  2 $\mu$m above the PACS grid. Even if this solution is technically possible to build, it would require a complete design study, which is not the goal of this work.  The solution \textit{B}, which also gives a high response efficiency, requires the insertion of a 51 $\mu$m thick silicon layer at a distance of 179 $\mu$m (airgap) above the bolometer grid. As the interpixels walls have a fixed height of  450 $\mu$m because of the etching process (Fig. \ref{Figure2}a), this solution would require a silicon waffle shaped structure encapsulated in the PACS array. Similar structures have been made in our laboratory, and were successfully mounted on existing PACS arrays. For technical and practical reasons, we first focused on a simpler solution, (\textit{C}), that gives a lower $C_{abs}$ coefficient (mainly because the response is not flat),  but which is technically easier to implement: a 138 $\mu$m thick silicon layer is deposited (glued) directly on the top of the 450 $\mu$m interpixel walls. Because of this simplicity of assembly, this solution has been selected to experimentally validate our method (see section 3).

\begin{figure}[htbp]
  \centering
  \includegraphics[width=8.5cm]{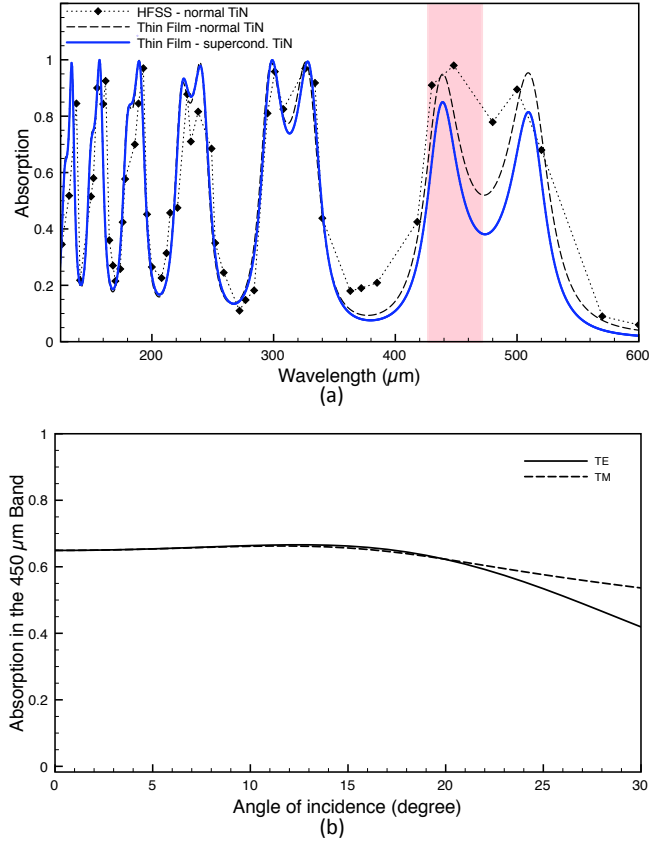}
  \caption{(Color online) (a) Simulated spectral response for the PACS + AR layer (solution \textit{C} on Fig. \ref{Figure6}a: airgap = 450 $\mu$m, silicon AR = 138 $\mu$m). The red surface represents the  450 $\mu$m atmospheric band. (b) Influence of the angle of incidence on the absorption coefficient in the 450 $\mu$m band.}
  \label{Figure7}
  \end{figure}

Once the 2D map is obtained by the model ($\sim$ 30 minutes of calculation time on a typical PC), the chosen solution is studied more in details (comparison with HFSS\texttrademark ).  Fig. \ref{Figure7}a shows the corresponding spectral response of solution \textit{C}. The average absorption coefficient is 88$\%$ in the 450 $\mu$m band with HFSS\texttrademark , 71\% with the matrix method for normal TiN and 65\% for superconducting TiN. The different curves are in very good agreement between 100 $\mu$m and 440 $\mu$m, but the matrix method shows a larger dip near 480 $\mu$m compared to the HFSS\texttrademark  case. The improvement in spectral response is  significative compared to the bare bolometer pixel (43$\%$ absorption). To use this detector in a scientific instrument, the parasitic peaks of absorption at shorter wavelengths (below 400 $\mu$m) need to be strongly attenuated with a combination of long-pass  and band-pass optical filters. The model has also been used to test the influence of the angle of incidence on the spectral response of the detector. Results on Fig. \ref{Figure7}b show that the absorption is constant up to 20$^{\circ}$ which is good enough for the requirements of a scientific bolometer camera (the final f-number is typically between 3 and 5 in a submillimeter camera with filled arrays, \cite{Talvard}).

\subsubsection{350-735 $\mu$m  Coverage}
Based on this method, we have investigated the possibility to cover all the atmospheric windows between 350 $\mu$m and 735 $\mu$m with a 20 $\mu$m $\lambda$/4 cavity PACS array topped with an adapted AR layer. The results corresponding to the best performance are presented in Table 2 and in Fig. \ref{Figure9}.

For each atmospheric band considered in this example, there is a particular airgap/AR combination that improves the spectral response. Except for the 350 $\mu$m band, the absorption is improved by a minimum factor 2 (6 in the 735 $\mu$m band). All these solutions require waffle shaped structures inserted in the original array.


\section{Experimental Results}

\subsection{Spectrometer Characterization}

We used a polarizing Fourier transform spectrometer (Martin-Puplett) to measure the spectral response of the bolometer arrays with or without an AR layer.  The experimental setup is detailed in Appendix B. In this setup, the global absorption of the $16 \times16$ pixels array is deduced from the reflected signal of the FTS beam on its surface. Because the incident beam is larger than one pixel and because it is not normal to the surface (the incident angle is $\sim$ 5$^{\circ}$), the interpixel silicon walls add a small scattering effect  which results in oscillations in the spectral response. The model has been modified in order to take this effect into account, and the result can be seen in Fig. \ref{Figure9}a for the normal PACS array case. Fig. \ref{Figure9}b shows the spectral response of the PACS+AR layer (solution \textit{C}). The experimental data look very similar to the model, with an average absorption of 74\% in the 450 $\mu$m band.

\subsection{Bolometer camera on a telescope}

We designed a camera for astrophysical observations in the 450 $\mu$m band equipped with a $16\times 16$ PACS+AR array, similar to the one characterized with the FTS. This instrument is called p-ArT\'eMiS \cite{Talvard}, and was installed at the Cassegrain focus of the APEX telescope in the Atacama desert in Chile. During the different runs (in 2007 and 2009), the detector showed the expected performances in terms of optics and sensitvity \cite{{Andre},{Minier}}.


\section{Conclusion}

Large arrays made of bolometers using  quarter wave cavities are now commonly used in submillimeter cameras for astrophysics. Once assembled, it is not easy to modify the intrinsic spectral properties of these detectors. However, there are cases where it is interesting to use an array that was originally designed for a particular band to operate in other bands. For this purpose, our group has developed a technique using a front end anti-reflecting dielectric layer that enhances the spectral response of bolometer arrays in other windows. To get the optimum parameters of the dielectric layer, we have developed a model based on the method of characteristic matrix for thin films. The results of the model have been compared to outputs from Ansoft HFSS\texttrademark and show very good accordance.
 
We applied this general method to the case of the Herschel PACS bolometer arrays which was developed for the 60 $\mu$m - 210 $\mu$m window.  The cavity of the PACS detectors is formed by the hybridization of two layers with 20 $\mu$m indium bumps. We showed that it is possible to have a good spectral response for all the atmospheric transmission windows between 350 $\mu$m and 735 $\mu$m by adding a dielectric layer (silicon) with a thickness calculated by the model, over the original array. This design has been tested and validated experimentally in a spectrometer, and used for science purpose in a small submillimeter camera. 





\section*{Appendix A: Details of the parameters for the square loops resonance.}

$H$ is defined by

\begin{equation}
H(p,s,\lambda)=\frac{p}{\lambda}\cos \theta 
\left[
\ln\left(\frac{1}{\sin\frac{\pi s}{2p}}\right)
+G(p,s,\lambda)
\right]
\end{equation}
where $\theta$ is the incident angle and

\begin{eqnarray*}
\lefteqn{G(p,s,\lambda) =}  \\
&& \frac{1}{2}\frac{(1-\phi^2)^2\left[ \left(1-\frac{\phi^2}{4}\right) (A_++A_-)+4\phi^2A_+A_- \right]}
{\left(1-\frac{\phi^2}{4}\right) +\phi^2\left(1+\frac{\phi^2}{2}-\frac{\phi^4}{8}\right) (A_++A_-)+2\phi^6A_+A_- }
\end{eqnarray*}
where

\begin{equation}
A_\pm=\frac{1}{\sqrt{1\pm\frac{2p\sin\theta}{\lambda}-\left(\frac{p\cos\theta}{\lambda}\right)^2}}-1 
\end{equation}
and

\begin{equation}
\phi=\frac{\sin\pi s}{2p}
\end{equation}

\section*{Appendix B: Martin-Puplett FTS}

Our spectrometer has a typical resolution of 0.25 cm$^{-1}$ in the 100 to 700 $\mu$m band and operates at cryogenic temperatures. Measurements were made in reflective mode by comparison with a perfectly reflecting mirror (see Fig. \ref{Figure10}). By considering that the power which is not reflected by the bolometer array is entirely absorbed, the spectral response is calculated with $A=1-R_{Array}/(C\times R_{Mirror})$, where $R_{Array}$ is the signal reflected from the bolometer array,  $R_{Mirror}$ is the signal reflected from the mirror and $C$ is a geometrical correcting factor.  The pixel response is obtained from the global response of the array.




\clearpage

\clearpage

\clearpage

\clearpage

\clearpage

\clearpage

\clearpage
  
\clearpage
  \begin{figure}[htbp]
  \centering
  \includegraphics[width=8.5cm]{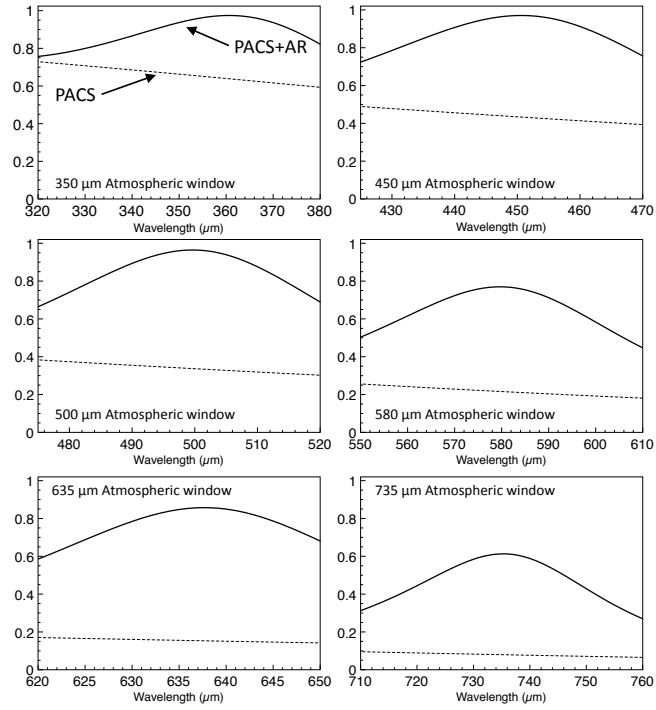}
  \caption{Spectral responses of a 20 $\mu$m $\lambda$/4 cavity covered by different  AR layers thicknesses and airgaps. Each curve corresponds to an atmospheric window defined in Table 2.}
  \label{Figure8}
  \end{figure}

\clearpage
\begin{figure}[htbp]
  \centering
  \includegraphics[width=8.5cm]{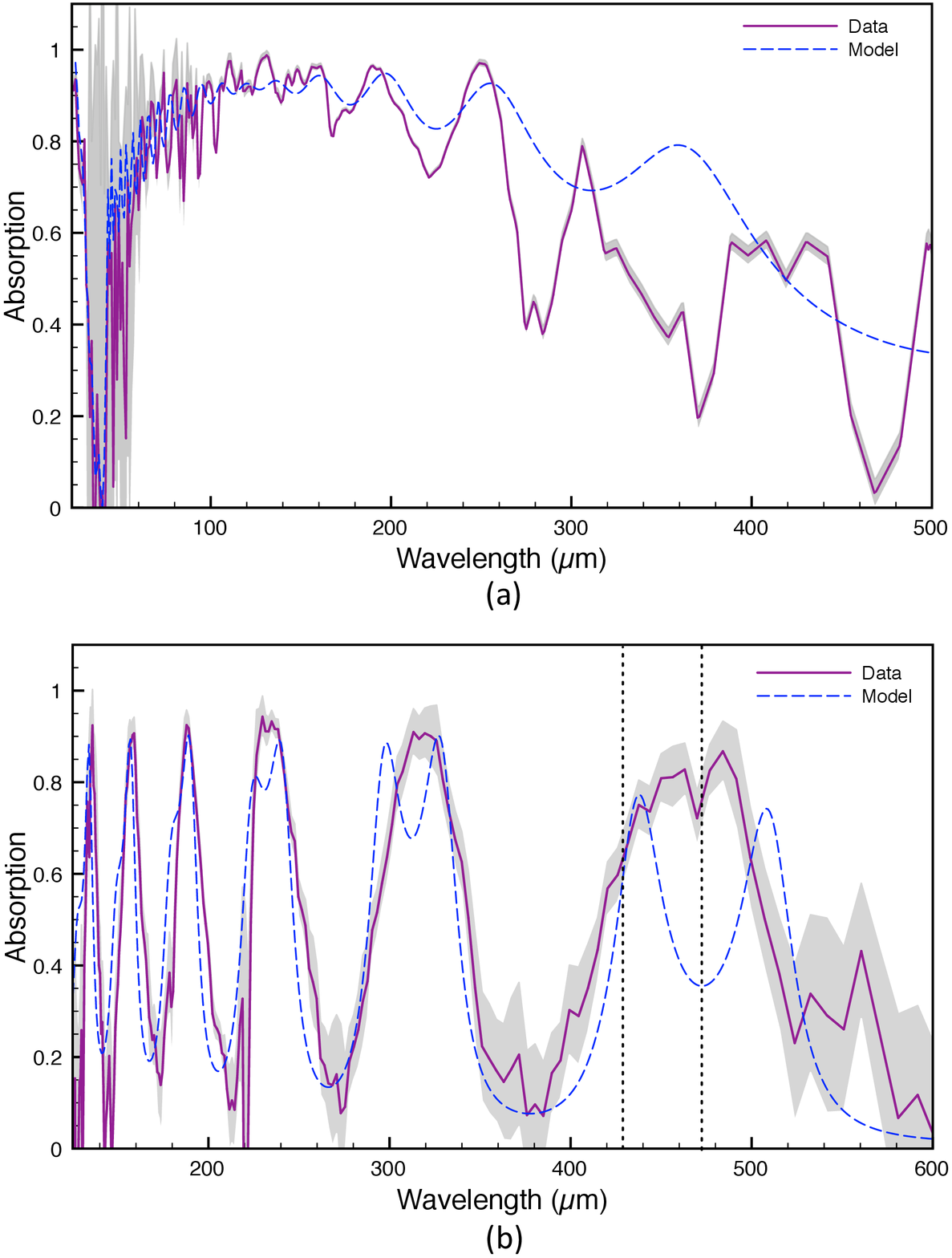}
  \caption{(Color online) (a) Experimental spectral response of the PACS array. The gray surface corresponds to the 3$\sigma$ error bars. (b) PACS+AR experimental spectral response (solution \textit{C} on Fig. \ref{Figure6}a, 450 $\mu$m window).}
  \label{Figure9}
  \end{figure}

\clearpage
 \begin{figure}[htbp]
  \centering
  \includegraphics[width=7.5cm, angle=270]{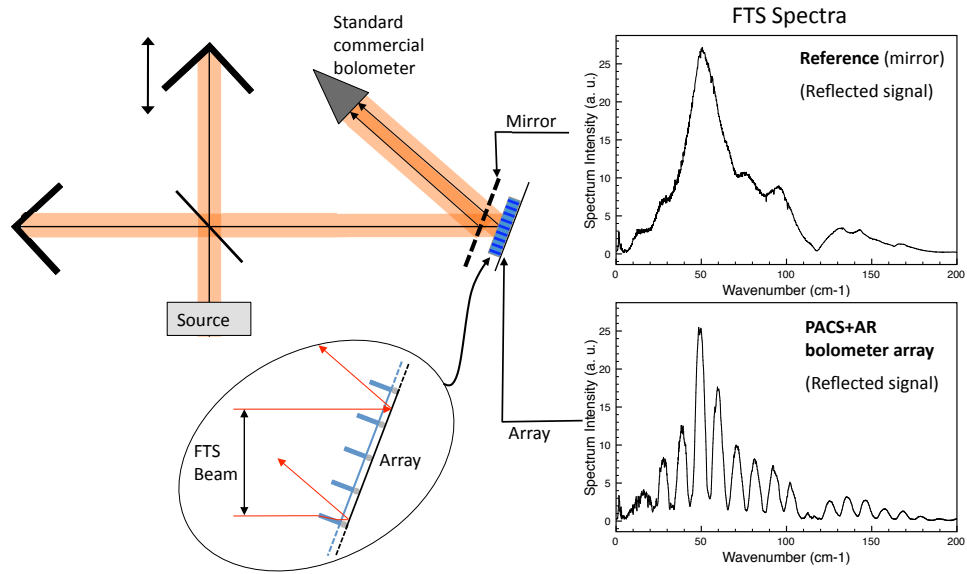}
  \caption{(Color online) Description of the Fourier Transform Spectrometer setup used for the spectral response characterization. The incident angle on the array has been exagerated for clarity (5$^{\circ}$ in reality). The two curves are examples showing typical spectra from reflected reference beam (above) and reflected array beam.}
  \label{Figure10}
  \end{figure}

\clearpage

\begin{table}[h] 
{\bf \caption{\label{table1}Layers parameters for PACS bolometers}}
\begin{center}
\begin{tabular}{p{1.5in}p{1.3in}p{1.4in}lp{2.0in}lp{1.5in}}\hline
Layer & Thickness & Refractive Index & Notes \\ \hline
reflector & 150 nm & - & reflective substrate \\
$\lambda/4$ cavity & 20 $\mu$m & 1 & vacuum cavity  \\
TiN absorber & 20 nm & $\tilde{n}_{abs}$ &  $\tilde{n}_{abs}=\sqrt{\frac{1}{2\rho \epsilon_0\omega}}(1-i)$ \\
Si grid & 5 $\mu$m & 2.2 & effective refractive index  \\
\hline
\end{tabular}
\end{center}
\end{table}

\clearpage
\begin{table}[h] 
{\bf \caption{\label{table2}Average absorption coefficient in different atmospheric bands}}
\begin{center}
\begin{tabular}{lccccc}\hline
Band & \multicolumn{1}{c}{PACS} & & \multicolumn{3}{c}{PACS+AR} \\
\cline{2-2}
\cline{4-6}
 & $Abs_x$ & & $Abs_x$ & Airgap ($\mu$m) & AR ($\mu$m)\\
\hline  
350 $\mu$m  & 68\%  & & 91\%  & 133 & 45 \\ 
450 $ \mu$m  & 43\%  & & 90.5\%  & 179 & 51 \\ 
500 $ \mu$m  & 36\%  & & 86\%  & 206 & 53 \\ 
580 $ \mu$m  & 23\%  & & 67\%  & 290 & 20 \\ 
635 $ \mu$m  & 16.5\%  & & 75\%  & 285 & 55 \\ 
735 $ \mu$m  & 8\%  & & 48\%  & 337 & 60 \\ 
\hline
\end{tabular}
\end{center}
\end{table}


\begin{thebibliography}{}

\bibitem{Pilbratt} G. L. Pilbratt, J. R. Riedinger, T. Passvogel, G. Crone, D. Doyle, U. Gageur, A. M. Heras, C. Jewell, L. Metcalfe, S. Ott, M. Schmidt, ``Herschel Space Observatory. An ESA facility for far-infrared and submillimetre astronomy,'' A\&A \textbf{518}, L1 (2010).

\bibitem{Nakagawa} T. Nakagawa, H. Murakami, ``Mid- and far-infrared astronomy mission SPICA,'' Advances in Space Research \textbf{40}, 679-683 (2007). 

\bibitem{Guesten} R. G\"usten, L. \AA. Nyman, P. Schilke, K. Menten, C. Cesarsky, R. Booth, ``The Atacama Pathfinder EXperiment (APEX) - a new submillimeter facility for southern skies -,'' A\&A \textbf{454}, L13-L16 (2006).

\bibitem{Sebring} T. Sebring, ``The Cornell Caltech Atacama Telescope: progress and plans 2010,'' \pspie \textbf{7733}, 77331X (2010).


\bibitem{Poglitsch} A. Poglitsch, C. Waelkens, N. Geis, H. Feuchtgruber, B. Vandenbussche, L. Rodriguez, O. Krause, E. Renotte, ``The Photodetector Array camera and Spectrometer (PACS) on the Herschel Space Observatory,'' A\&A \textbf{518}, L2 (2010).

\bibitem{Billot06} N. Billot, P. Agn\`ese, J.-L. Augu\`eres, A.  B\' eguin, A. Bou\`ere, O. Boulade, C. Cara, C. Clou\'e, E. Doumayrou, L. Duband, B. Horeau, I. le Mer, J. Lepennec, J. Martignac, K. Okumura, 
V. Rev\'eret, M. Sauvage, F. Simoens, L. Vigroux, 
``The Herschel/PACS 2560 bolometers imaging camera,'' \pspie \textbf{6265} 62650D (2006).

\bibitem{Richards94} P. L. Richards, ``Bolometers for infrared and millimeter waves,'' J. Appl. Phys. \textbf{76}, 1-24 (1994).

\bibitem{Griffin02} M. J. Griffin, J. J. Bock, W. K. Gear, ``Relative performance of filled and feedhorn-coupled focal-plane architectures,'' \ao \textbf{41}, 6543-6554 (2002).

\bibitem{Agnese99} P. Agn\`ese, C. Buzzi, P. Rey, L. Rodriguez, J.-L. Tissot, ``New technological development for far-infrared bolometer arrays,'' \pspie \textbf{3698}, 284 (1999). 

\bibitem{Hadley47} L. N. Hadley, D. M. Dennison, ``Reflection and transmission interference filters,'' \josa \textbf{37}, 451-465 (1947). 

\bibitem{Salisbury} W. W. Salisbury, ``Absorbent body for electromagnetic waves,'' US patent 2599944, June 10 1952.

\bibitem{Fante88} R. L. Fante, M. T. McCormack, ``Reflection properties of the Salisbury screen,'' \aprop \textbf{36}, 1443-1454 (1988).

\bibitem{Zhang93} J. Zhang, W. Cui, M. Juda, D. McCammon, R. L. Kelley, S. H. Moseley, C. K. Stahle, A. E. Szymkowiak, ``Hopping conduction in partially compensated doped silicon,'' \prb \textbf{48}, 2312-2319 (1993).

\bibitem{Bock95} J. J. Bock, D. Chen, P. D. Mauskopf, A. E. Lange, ``A novel bolometer for infrared and millimeter-wave astrophysics,'' Space Sci. Rev. \textbf{74}, 229-235 (1995).

\bibitem{Mirotznik} M. Mirotznik, W. A. Beck, D. Prather, R. Vollmerhausen, R. Driggers, ``Optical absorption modeling of thermal infrared detectors by use of the finite-difference time-domain method,'' \ol \textbf{26}, 280-282 (2001).

\bibitem{Abeles} F. Abel\`es, ``La th\'eorie g\'en\'erale des couches minces,''  Le Journal de Physique et le Radium,  \textbf{11}, 307-310 (1950).

\bibitem{Macleod} H.A. Macleod, \textit{Thin-Film Optical Filters: 3rd edition} (Adam Hilger Ltd. Bristol, 1986). 

\bibitem{Matula} R. A. Matula, ``Electrical resistivity of copper, gold, palladium, and silver,'' J.  Phys. Chem. Ref. Data \textbf{8}, 1147-1298 (1979).
 
\bibitem{Langley} R. J. Langley, E. A. Parker, ``Equivalent circuit model for arrays of square loops,'' Electronics Letters \textbf{18}, 294-296 (1982).

\bibitem{Monacelli} B. Monacelli, J. B. Pryor, B. A. Munk, D. Kotter, G. D. Boreman, ``Infrared frequency selective surface based on circuit-analog square loop design,''  \aprop \textbf{53}, 745-752 (2005).

\bibitem{BCS} J. Bardeen, L. N. Cooper, J. R. Schrieffer, ``Theory of superconductivity,'' Phys. Rev. \textbf{108}, 1175--1204 (1957).

\bibitem{Mattis} D. C. Mattis, J. Bardeen, ``Theory of the anomalous skin effect in normal and superconducting metals,'' Phys. Rev. {111}, 412-417 (1958).

\bibitem{Kautz} R. L. Kautz, ``Picosecond pulses on superconducting striplines,'' J. Appl. Phys. \textbf{49}, 308-314 (1978).

\bibitem{Hilsum} C. Hilsum, ``Infrared absorption of thin metal films,''  \josa \textbf{44}, 188-191 (1954).

\bibitem{Motamedi} M. E. Motamedi, W. H. Southwell, W. J. Gunning, ``Antireflection surfaces in silicon using binary optics technology,'' \ao \textbf{31}, 4371-4376 (1992).

\bibitem{Talvard} M. Talvard, Ph Andr\'e, L. Rodriguez, J. Le Pennec, C. De Breuck, V. Rev\'eret, P. Agn\`ese, O. Boulade, E. Doumayrou, D. Dubreuil, E. Ercolani, P. Gallais, B. Horeau, P.-O. Lagage, B. Leriche, M. Lortholary, J. Martignac, V. Minier, E. Pantin, D. Rabanus, J. Relland, G. Willmann, ``Recent results obtained on the APEX 12m antenna with the ArTeMiS prototype camera,'' \pspie \textbf{7020}, 702009.1-702009.9 (2008).

\bibitem{Andre} Ph. Andr\'e, V. Minier, P. Gallais, V. Rev\'eret, J. Le Pennec, L. Rodriguez, O. Boulade, E. Doumayrou, D. Dubreuil, M. Lortholary, J. Martignac, M. Talvard, C. De Breuck, G. Hamon, N. Schneider, S. Bontemps, P.-O. Lagage, E. Pantin, H. Roussel, M. Miller, C. R. Purcell, T. Hill, J. Stutzki, ``First 450 $\mu$m dust continuum mapping of the massive star-forming region NGC 3576 with the P-ArT\'eMiS bolometer camera,'' Astron. Astrophys. \textbf{490}, L27-L30 (2008).

\bibitem{Minier} V. Minier, Ph. Andr\'e, P. Bergman, F. Motte, F. Wyrowski, J. Le Pennec, L. Rodriguez, O. Boulade, E. Doumayrou, D. Dubreuil, P. Gallais, G. Hamon, P.-O. Lagage, M. Lortholary, J. Martignac, V. Rev\'eret, H. Roussel, M. Talvard, G. Willmann, H. Olofsson, ``Evidence of triggered star formation in G327.3-0.6. Dust-continuum mapping of an infrared dark cloud with P-ArT\'eMiS,'' Astron. Astrophys. \textbf{501}, L1-L4 (2009).



\end{thebibliography}
\end{document}